\documentclass[citeauthoryear]{apinv}
\usepackage{epsfig,cite,graphics}
\usepackage{marvosym}
\usepackage[utf8]{inputenc}
\usepackage[pagewise]{lineno}
\usepackage{graphicx}
\usepackage{appendix}
\usepackage{verbatim}
\usepackage{tikz}
\usepackage{placeins}
\usepackage{amsmath}
\usepackage{soul}
\usepackage{amssymb}

\begin{document}
\nolinenumbers

\title{Velocity-field estimation techniques for on-disk solar coronal bright fronts (CBFs) and shock-wave propagation}
\titlerunning{Velocity-field estimation for CBFs}
\author{Oleg Stepanyuk\inst{1}, Kamen Kozarev\inst{1}}
\authorrunning{O. Stepanyuk, K. Kozarev}
\tocauthor{Oleg Stepanyuk, Kamen Kozarev}
\institute{Institute of Astronomy and National Astronomical Observatory, Bulgarian Academy of Sciences, Sofia 1784, Bulgaria \newline \email{stepanyuk.oleg@aol.com}}
\papertype{Submitted on 15.07.2025; Accepted on 19.08.2026}
\maketitle

\begin{abstract}
Coronal mass ejections (CMEs) are large expulsions of plasma and magnetic fields from the solar corona into the heliosphere. Coronal shock waves are driven by CMEs, and accurate Solar Energetic Particles (SEP) forecasting requires knowing when and where shocks form, how fast they propagate, and how they interact with the coronal magnetic field. By measuring the speed, geometry and temporal evolution of coronal bright fronts (CBFs), one can create more realistic predictions of SEP onset and intensity at Earth, thus velocity-field estimation techniques supply these critical parameters.
High-resolution velocity field maps can provide even deeper insight. Instead of yielding only an overall speed of the front, techniques like FLCT and optical flow can show the spatial distribution of velocities across the entire wave dome or front. These analyses reveal which parts of the front are moving fastest and how the front might be deforming as it expands. 
Here we discuss a set of practical parameters for Lucas--Kanade, Horn--Schunck, FLCT, and Farneb{\"a}ck algorithms, arrived at by combining analytic reasoning with empirical measurements on a single event---and demonstrate the potential of these techniques for on-disk shock-wave propagation during the X1.5-class event that took place 10 May 2022, processed by the wavelet-based Wavetrack technique. A better understanding of how to fine-tune these parameters could improve the precision of algorithms running in pipelines, and is especially valuable where per-event hand-tuning is impractical.

\end{abstract}

\section*{Introduction}
\label{introduction}
The motivation for studying coronal shock waves is tightly linked to space weather: shocks driven by CMEs are a primary source of solar energetic particles (SEPs), high-energy protons and ions that can pose radiation hazards to spacecraft and astronauts. Reliable SEP forecasting requires understanding when and where particles are accelerated in a solar eruption and how they travel to Earth. Velocity field estimation techniques contribute to this in several ways.
First, identifying a CBF as a shock and measuring its speed, geometry, and evolution provides the initial conditions for particle acceleration models. The efficiency of particle acceleration at a shock depends on the shock's speed and angle relative to the magnetic field (quasi-parallel vs quasi-perpendicular), among other factors \cite{Kozarev_2017_25}.
Secondly, understanding the propagation of SEPs through the corona and interplanetary space benefits from knowing the shock propagation history. SEPs released by a moving shock will travel along the interplanetary magnetic field to reach Earth or other spacecraft. The timing of SEP onset and their intensity can be better predicted if we know when the shock formed and how fast it traveled through the corona. Velocity maps, combined with coronagraph data in the higher corona (e.g., from SOHO/LASCO or STEREO), allow the construction of a continuous height--time profile of the front.
With higher cadence data from STEREO and SDO, two-phase kinematics can be observed: an initial fast phase (several hundred km/s) that often decelerates over time \cite{Long:2011}.
High-resolution velocity field maps have provided even deeper insight. Instead of yielding only an overall speed of the front, techniques like FLCT and optical flow can show the spatial distribution of velocities across the entire wave dome or front. These analyses reveal, for instance, which parts of the front are moving fastest and how the front might be deforming as it expands. Often, the flanks of the coronal wave (the edges sweeping to the north/south of the eruption center) are observed to move faster than the wave segment directly above the eruption or on the disk.

Image-by-image segmentation, traditionally performed using algorithmic methods and more recently explored with data-driven approaches, often considers only the spatial aspects of image features, entirely overlooking their temporal evolution. Beyond their value in enhancing the qualitative understanding of physical processes, velocity field estimation techniques can further help differentiate genuine details from spurious features as these methods operate within a space-time context rather than being limited to purely spatial processing. Simply put, capturing feature movement helps to distinguish moving details from static ones, even if they have similar intensities or shapes.\cite{Stepanyuk:2025}
The velocity field of a global EUV wave observed by SDO/AIA on 2021-05-07, derived using local correlation tracking, lets us measure instantaneous velocities of different parts of the CBF and how they change over time. In the 2021-05-07 event shown, the EUV wave's flanks reached $\sim$500\,km/s early on, whereas later frames showed those flanks slowing down and the northern part of the front dropping below 200\,km/s \cite{Stepanyuk2024_25}. This kind of information is critical for understanding the wave's evolution and energy dissipation. Velocity fields can be used to derive physical parameters of the corona: combining an observed wave speed with assumptions of fast-mode wave behavior yields an estimate of the local magnetosonic speed, and hence the coronal magnetic field strength. In the 2010 event \cite{Long:2011}, the range of observed CBF speeds (380--460\,km/s) implied a quiet coronal magnetic field of order 1--2 Gauss in the low corona. Thus, velocity estimation techniques are not just about kinematics -- they also provide a diagnostic tool for the coronal environment (plasma density and magnetism) through which the wave propagates.

In solar physics, the FLCT method is often considered a first-choice technique for determining horizontal flows, particularly in active regions. The method calculates local shifts between two images in the Fourier domain, which is particularly useful for tracking solar granulation and magnetic fields. By computing cross-correlations in Fourier space, it can be efficient and able to identify local displacements effectively \cite{Fisher:2008}.

\cite{Byrne:2013} addresses the FLCT sampling-cadence problem within the framework of numerical differentiation, emphasizing that naive implementations of basic difference formulas are inadequate in scenarios with infrequent or irregular data sampling. The fundamental difficulty arises from errors introduced in differentiation processes due to non-uniform or sparse sampling intervals. Consequently, more sophisticated methods, such as higher-order polynomial fits or adaptive numerical schemes, become necessary in these situations. This challenge prompted our exploration of alternative velocity field estimation techniques \cite{Stepanyuk:2024}, where we mainly used Horn--Schunck \cite{horn:1981} and Lucas--Kanade \cite{lucas:1981} velocity estimation methods as an alternative for FLCT. Here we extend this research by performing a comparative analysis of velocity field estimation techniques for on-disk wave propagation.

\section{Method}
\label{method}

In the context of estimating velocity fields from temporally separated frames, Fourier Local Correlation Tracking (FLCT) and classical optical-flow algorithms such as Lucas--Kanade, Horn--Schunck an Farneb{\"a}ckprovide distinct methodological approaches, each with tunable parameters that significantly influence their sensitivity to different motion scales and structures. This section first introduces the formal description of each algorithm (Sec.~\ref{sec:algorithms}). The remainder of the section then works through a set of parameters for detecting moderately sized circular features, combining \emph{analytic} reasoning with \emph{empirical} statistics measured directly on the AIA 193\,\AA{} running-difference frames of the X1.5-class event of 10 May 2022 (Sec.~\ref{sec:fitting}). This is presented as a worked example of how such reasoning can give an algorithm a sensible starting point rather than being tuned blind; Sec.~\ref{sec:demonstration} then demonstrates the resulting algorithm performance on the same event.
The 10 May 2022 event therefore serves as the running case study throughout: it provides the calibration data for the parameters and the demonstration case for the resulting algorithms.
\subsection*{Algorithms}
\label{sec:algorithms}

\subsubsection*{Fourier Local Correlation Tracking (FLCT)}

FLCT estimates the velocity field by calculating the local cross-correlation of image patches in the Fourier domain. It is designed to find the displacement that best aligns local regions between two images. The core idea is to find the shift $(\delta x, \delta y)$ that maximizes the local cross-correlation function $C(\delta x, \delta y)$ between image patches. Let $f_1(x, y)$ and $f_2(x, y)$ be two consecutive image frames. The local correlation function is computed as

\begin{equation}
C(\delta x, \delta y) = \mathcal{F}^{-1} \left\{ \mathcal{F}[G(x, y) f_1(x, y)] \cdot \overline{\mathcal{F}[G(x, y) f_2(x, y)]} \right\}(\delta x, \delta y),
\label{eq:flct}
\end{equation}

where $G(x, y)$ is a Gaussian window, $\mathcal{F}$ denotes the Fourier transform, and the overline denotes the complex conjugate.
The velocity vector is then estimated from the peak of the correlation function in each windowed region. The technique therefore exposes three control parameters: the correlation patch size $W_\mathrm{FLCT}$, the Gaussian weighting width $\sigma_G$, and an acceptance threshold $\tau$ on the peak correlation that filters out weak detections.

\subsubsection*{The Horn--Schunck method}

The Horn--Schunck method is a classical approach to estimating optical flow. It is based on the assumption that the apparent velocity of brightness patterns in an image sequence varies smoothly almost everywhere. The key assumptions are brightness constancy (the intensity of a moving object remains constant between frames) and a smoothness constraint (the optical flow field varies smoothly over the image). Let $I(x, y, t)$ represent the image intensity at position $(x, y)$ and time $t$. Assuming small motion between frames, the brightness constancy constraint reads
\begin{equation}
\frac{\partial I}{\partial x} u + \frac{\partial I}{\partial y} v + \frac{\partial I}{\partial t} = 0,
\end{equation}
where $u$ and $v$ are the horizontal and vertical components of the optical flow. To make the problem well-posed, Horn and Schunck added a regularization term that enforces smoothness:
\begin{equation}
E(u, v) = \iint \left[ \left( I_x u + I_y v + I_t \right)^2 + \alpha^2 \left( \|\nabla u\|^2 + \|\nabla v\|^2 \right) \right] dx\,dy,
\label{eq:hs}
\end{equation}
where $\alpha$ is a regularization parameter controlling smoothness and $\nabla u$, $\nabla v$ are the gradients of the flow field. The Euler--Lagrange equations are used to minimize the energy functional; this leads to an iterative Jacobi update scheme where each flow component is refined using local averages and gradient information. The two control parameters that the user must select are $\alpha$ and the number of Jacobi iterations $N_\mathrm{it}$.

\subsubsection*{Lucas--Kanade optical flow}

The Lucas--Kanade method assumes that the optical flow is constant within a small window of side $W$. It solves the optical flow equation
\begin{equation}
I_x u + I_y v + I_t = 0
\end{equation}
by performing a least-squares fit in this local region. Minimizing the squared residual over a window of pixels leads to the normal equations
\begin{equation}
\begin{bmatrix}
\sum I_x^2 & \sum I_x I_y \\
\sum I_x I_y & \sum I_y^2
\end{bmatrix}
\begin{bmatrix}
u \\ v
\end{bmatrix}
=
-
\begin{bmatrix}
\sum I_x I_t \\
\sum I_y I_t
\end{bmatrix},
\label{eq:lk}
\end{equation}
where the sums run over all pixels in the local window. The system is well-conditioned only where the structure tensor on the left-hand side has both eigenvalues above some threshold $\lambda_\mathrm{min}$. The spatial derivatives $I_x, I_y$ are evaluated with a Sobel kernel of size $k$. The three control parameters are therefore $(W, k, \lambda_\mathrm{min})$.

\subsubsection*{Farneb{\"a}ck: polynomial expansion of neighborhoods}

The Farneb{\"a}ck optical-flow algorithm \cite{Farneback:2002} is a dense optical-flow method that estimates motion between two images by modeling pixel neighborhoods as polynomial expansions. Unlike sparse methods (e.g., Lucas--Kanade), which track specific features, Farneb{\"a}ck's algorithm computes flow vectors for every pixel, offering a full motion field. The core idea is to approximate the local neighborhood of each pixel using a quadratic polynomial, allowing a smooth representation of image intensity variations. Specifically, the algorithm models an image patch around a pixel $\mathbf{x}$ as
\begin{equation}
f(\mathbf{x}) \approx \mathbf{x}^T \mathbf{A} \mathbf{x} + \mathbf{b}^T \mathbf{x} + c,
\label{eq:farneback}
\end{equation}
where $\mathbf{A}$ is a symmetric matrix representing the second-order terms, $\mathbf{b}$ is a vector for the linear terms, and $c$ is a constant. This polynomial expansion is computed for both frames; the motion is estimated by comparing the polynomial models. The method captures both large and small displacements through a multi-level pyramid. The final output is a dense flow field $\mathbf{v}(x, y) = (u(x, y), v(x, y))$. The user-selectable parameters are the polynomial neighborhood size $n$ (\texttt{poly\_n}), the Gaussian weighting $\sigma$ used to compute the polynomial coefficients (\texttt{poly\_sigma}), the averaging window $W_F$ (\texttt{winsize}), the pyramid scale $s$ (\texttt{pyr\_scale}) and the number of refinement iterations.


\section{Parameter fitting: analytic principles and AIA-based calibration}
\label{sec:fitting}

The parameter values below are calibrated on the X1.5-class flare that began at 13:50\,UT on 10 May 2022, originating from NOAA active region AR\,13006, located near the center of the solar disk as seen from Earth. The event triggered significant ionospheric disturbances, evident in variations of Total Electron Content (TEC) measured using GNSS data \cite{Lopez-Urias:2023}, and was the subject of a detailed shock-wave analysis by \cite{Rigney:2024}, who tracked the propagation of a large CBF through a coronal channel in the low solar atmosphere and correlated its movement with the production of radio bursts. Beyond its scientific interest, the event was chosen for its processing complexity, which makes it valuable for fine-tuning algorithm setups and parameters. Throughout this section we use ten Level-1 AIA 193\,\AA{} frames from 13:48 to 13:57\,UT to measure the empirical inputs that drive the parameter fitting.

Rather than a rigorous derivation, what follows is a worked example: for each algorithm we combine a simple analytic (scale-matching) argument with a handful of quantities measured directly on the AIA running-difference frames of the 10 May 2022 event, to arrive at a practical starting point for the four algorithms' control parameters. These relations give parameter estimates rather than unique optimal values, and the point of walking through them is methodological -- to show that this kind of reasoning can save an algorithm from being tuned blind, and give it something sensible to start from -- rather than to establish a definitive configuration for every dataset. We summarize the empirical inputs measured on AIA first, then walk through each algorithm in turn.

\subsection{Empirical inputs measured from AIA 193\,\AA{}}
\begin{figure}[!htb]
  \begin{center}
    \centering{\epsfig{file=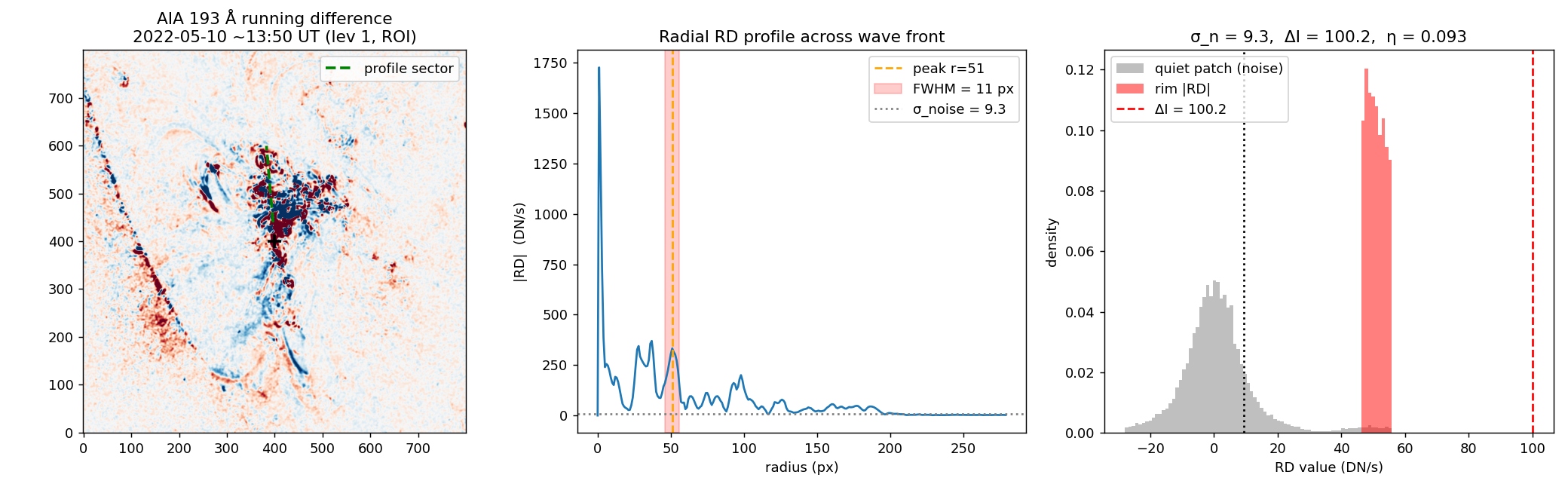, width=1.0\textwidth}}
    \caption{Empirical statistics measured on AIA 193\,\AA{} running-difference frames of the 10 May 2022 X1.5 event ($\sim$13:50\,UT). \textit{Left:} running-difference ROI with the strong-rim azimuthal sector (dashed green) along which the radial profile is sampled. \textit{Middle:} radial profile of $|\mathrm{RD}|$ across the wave front; the rim FWHM is $w = 11$\,px (orange band), well inside the 10--30\,px target band. The horizontal dotted line marks the quiet-corona noise floor $\sigma_n = 9.3\,\mathrm{DN\,s^{-1}}$ (1.4826$\,\times\,$MAD). \textit{Right:} histograms of $|\mathrm{RD}|$ for the quiet patch (gray) and the rim pixels (red); the rim contrast $\Delta I = 100.2\,\mathrm{DN\,s^{-1}}$ (75th percentile of rim $|\mathrm{RD}|$) and the dimensionless noise/signal ratio $\eta = \sigma_n/\Delta I = 0.093$ follow directly.}
    \label{fig:diagnostics}
  \end{center}
\end{figure}

Running-difference cubes $\mathrm{RD}[t] = I[t+1] - I[t]$ are formed in raw $\mathrm{DN\,s^{-1}}$ (each frame normalized by FITS EXPTIME for photometric consistency, which stores the actual shutter open time of that frame (in seconds)). The eruption centre is identified as the brightest RD pixel near AR\,13006; a quiet-corona patch with the lowest RD standard deviation provides a noise reference. The strongest-rim azimuthal sector (southern flank) yields the radial profile of the wave rim. Figure~\ref{fig:diagnostics} illustrates the three measurements that drive the subsequent parameter fitting.

These measurements give: an AIA 193\,\AA{} plate scale $p = 435$\,km\,px$^{-1}$; an AIA cadence $\Delta t = 12$\,s; a wave-rim FWHM (southern flank) $w = 11$\,px; a rim contrast (75th percentile of rim $|\mathrm{RD}|$) $\Delta I \approx 100$\,DN\,s$^{-1}$; a quiet-corona noise level (1.4826$\,\times\,$MAD) $\sigma_n \approx 9$\,DN\,s$^{-1}$, giving a dimensionless noise-to-signal ratio $\eta = \sigma_n/\Delta I \approx 0.09$; and a rim gradient magnitude (75th percentile) $\|\nabla I\| \approx 337$\,DN\,s$^{-1}$\,px$^{-1}$. Combined with the wave-speed range of 500--1000\,km\,s$^{-1}$ reported for this event by \cite{Rigney:2024}, these imply an inter-frame displacement budget of $\Delta = v\Delta t/p \approx 13.8$--27.6\,px.

The measured rim FWHM $w \approx 11$\,px falls inside the 10--30\,px target band declared above, which is at least internally consistent with that choice. Each of the recommended values below draws on a subset of $\{w, \Delta I, \sigma_n, \eta, \|\nabla I\|, \Delta\}$ together with a simple analytic scale-matching relation.

\subsection{Parameter choices: a worked example}

\paragraph{Scale-matching principles.} Each of the parameter choices below reduces to matching one of two algorithm scales to one of two data scales, as an example of the kind of reasoning that can give a starting point rather than a blind guess:
\begin{enumerate}
\item \textbf{Spatial scale.} The algorithm's intrinsic spatial parameter (window, kernel, $\sigma$, smoothing length) must bracket the rim half-width $w/2$ (to resolve the edge) and stay below the feature radius $r = L/2$ (to remain coherent across the feature).
\item \textbf{Displacement scale.} Any parameter that controls the maximum trackable displacement (LK pyramid level, Farneb{\"a}ck $\mathrm{winsize}\times\mathrm{pyr\_scale}$, FLCT window) must absorb $\Delta = v\Delta t/p$, the per-frame pixel shift of the wave.
\end{enumerate}

\subsubsection*{Lucas--Kanade: window, kernel, eigenvalue threshold}

\paragraph{Window size $W$.} The LK aperture must hold one edge of the rim
while keeping $\vec{v}$ approximately constant inside the window. We use
$W \gtrsim w/2$ as a practical criterion for resolving the observed rim,
while $W \lesssim r$ is used as a practical criterion for retaining local
flow coherence across the feature. With $w = 11$~px, the first criterion
gives $W \gtrsim 5.5$~px. We therefore select the smallest odd window
satisfying this condition, $W = 7$.

\paragraph{Sobel kernel $k$.} The Sobel kernel must sample the rim profile without blurring it across more than one rim width. Setting $k \approx w/2$, with $w = 11$\,px, and rounding to the nearest odd integer gives $k = 5$.

\paragraph{Eigenvalue threshold $\lambda_{\min}$.}
For the normalized image $I_{\rm norm}=I/S$, the characteristic gradient across a rim of width $w$
and intensity contrast $\Delta I$ is estimated as
\begin{equation}
|\nabla I_{\rm norm}|
\sim
\frac{\Delta I/S}{w}.
\label{eq:grad_norm}
\end{equation}
This width-averaged estimate is a deliberately conservative proxy for the local gradient: it
spreads the total rim contrast $\Delta I$ over the full rim width $w$, rather than using the
sharper pointwise gradient $\|\nabla I\|\approx337$~DN~s$^{-1}$~px$^{-1}$ measured directly above.
The conservative, width-averaged proxy yields a smaller, safer threshold that is
less sensitive to the single steepest pixel along the rim and reduces the risk of rejecting
valid, moderately sloped windows.
The corresponding characteristic scale of the structure-tensor
eigenvalues is therefore
\begin{equation}
\lambda_{\rm char}
\sim
W^2
\left(
\frac{\Delta I/S}{w}
\right)^2.
\label{eq:lambda_char}
\end{equation}
For $W=7$, $\Delta I=100$~DN~s$^{-1}$,
$w=11$~px, and $S=10^4$~DN~s$^{-1}$, Eq.~\eqref{eq:lambda_char} gives
\begin{equation}
\lambda_{\rm char}
\sim
7^2
\left(
\frac{100/10^4}{11}
\right)^2
\simeq4.05\times10^{-5}.
\end{equation}
We therefore use $\lambda_{\min}=5\times10^{-5}$ as a practical
threshold for rejecting poorly conditioned Lucas--Kanade windows.
The estimate in Eq.~\eqref{eq:lambda_char} is a characteristic scale for threshold
selection rather than an exact prediction of the smaller
structure-tensor eigenvalue.

\subsubsection*{Farneb{\"a}ck: polynomial size, smoothing, pyramid}

\paragraph{Polynomial neighborhood $n$.} The polynomial model in Eq.~\eqref{eq:farneback} must span at least the rim half-width, requiring $n-1 \gtrsim w/2$. With $w = 11$\,px the smallest odd integer satisfying this rule is $n = 7$.

\paragraph{Polynomial smoothing $\sigma$.} We adopt the following practical relation between the polynomial neighborhood size and its associated Gaussian smoothing,
\begin{equation}
\sigma \simeq 0.3(n-1)-0.8,
\end{equation}
which gives $\sigma=1.0$ for $n=7$. We therefore use $n=7$ and
$\sigma=1.0$ for the polynomial expansion. This relation is used here
as a practical parameter-selection rule rather than as a formal
requirement of the Farneb{\"a}ck method, and it involves no AIA quantity.

\paragraph{Averaging window $W_F$.} The averaging window must encompass at least one rim width and, after pyramid downsampling, must be able to absorb the maximum displacement $\Delta_\mathrm{max}$. With $w = 11$\,px and $\Delta_\mathrm{max} \approx 27.6$\,px, an odd window of $W_F = 15$ satisfies both constraints at the base level.

\paragraph{Pyramid scale $s$.} The pyramid must satisfy $W_F\, s^{-L} \gtrsim \Delta_\mathrm{max}$ over $L$ levels. With $W_F = 15$ and $\Delta_\mathrm{max} \approx 27.6$\,px, a single coarse level $L=1$ at $s = 0.3$ already gives an effective window of $\sim 50$\,px, which as a scale-matching estimate appears large enough to cover the displacement budget.

\paragraph{Iterations.} Gauss--Seidel updates of the displacement field converge to within $\sim 1\%$ in $\approx 10$ sweeps for the structure-tensor conditioning typical of our problem. No AIA quantity enters this estimate.

\subsubsection*{FLCT: window, weighting, threshold}

\paragraph{Correlation window $W_\mathrm{FLCT}$.} The window must cover the feature on its largest scale; for a Gaussian profile of FWHM equal to the feature size $L_\mathrm{max}$ one requires $W \gtrsim 1.5\, L_\mathrm{max}/2.355$. For $L_\mathrm{max} \approx 30$\,px (the upper bound of the target range), this gives $W \gtrsim 19.1$; the smallest odd window satisfying this bound is $W = 21$. This bound is set by the chosen target scale and does not consume an AIA-measured quantity.

\paragraph{Gaussian weighting.} The Gaussian apodization $G(x,y)$ in Eq.~\eqref{eq:flct} is mandatory to suppress wrap-around artefacts in the FFT-based correlation; without it spectral leakage spreads the correlation peak and biases the displacement.

\paragraph{Correlation threshold $\tau$.}  For Gaussian noise of relative amplitude $\eta = \sigma_n/\Delta I$, the expected correlation peak is $\langle C\rangle \approx 1/(1+\eta^2)$. With $\eta = 0.09$ this predicts $\tau \approx 0.99$ for AIA alone. The value $\tau = 0.3$ adopted here is deliberately set instead by the \emph{worst-instrument floor} in our wider processing pipeline (LASCO C2/C3, where $\eta \sim 1$), so that the same threshold can be applied across instruments without prejudicing AIA detections.

\subsubsection*{Horn--Schunck: regularization, iterations}


\paragraph{Regularization $\alpha$.}
Regularization $\alpha$. The Horn--Schunck regularization controls the degree of spatial smoothing of the estimated flow. As a heuristic scale estimate, we write the corresponding smoothing length as
\begin{equation}
\ell_{\rm HS} \sim
\frac{\alpha}{\|\nabla I\|/S},
\end{equation}
where $S \approx 1 \times 10^4$ DN s$^{-1}$ is the intensity scale used for the normalization. Using the measured gradient magnitude $\|\nabla I\| = 337$ DN s$^{-1}$ px$^{-1}$ gives
\begin{equation}
\frac{\|\nabla I\|}{S}
=
0.0337\ {\rm px}^{-1}.
\end{equation}
Requiring the characteristic smoothing scale to be of order $15$--$22$ px therefore gives the corresponding heuristic range
\begin{equation}
\alpha \sim
\ell_{\rm HS}\frac{\|\nabla I\|}{S}
\approx 0.51\text{--}0.74.
\end{equation}
This estimate is used only to constrain a sensible parameter range; it is not an exact prediction of the effective smoothing length of the discrete Horn--Schunck solver, which also depends on the discretization and iteration scheme. We therefore adopt the somewhat broader practical range $\alpha = 0.3$--$1.0$ for the calculations below.

\paragraph{Iterations $N_{\rm it}$}. The number of Jacobi iterations is not determined uniquely by the spatial scale argument above, because convergence depends on the regularization strength, the local image gradients, the discretization, and the initial flow field. We therefore do not interpret the iteration count as an analytically derived physical scale. Instead, we use $N_{\rm it}=500$ as a conservative fixed iteration budget for the present processing, chosen to allow the iterative solution to approach a stable flow field while keeping the computational cost practical.

The recommended values arrived at above are, for convenience: Lucas--Kanade window $W=7$, Sobel kernel $k=5$, eigenvalue threshold $\lambda_\mathrm{min}=5\times10^{-5}$; Farneb{\"a}ck pyramid scale (\texttt{pyr\_scale}) $=0.3$, window size (\texttt{winsize}) $=15$, $10$ iterations, polynomial neighborhood (\texttt{poly\_n}) $=7$, polynomial smoothing (\texttt{poly\_sigma}) $=1.0$; FLCT correlation window size $=21$, correlation threshold $\tau=0.3$, with Gaussian weighting applied; Horn--Schunck regularization $\alpha=0.3$--$1.0$ and $500$ iterations.

\bigskip
The only quantities measured directly on the AIA data throughout this section are the handful listed above (plate scale, cadence, rim width, contrast, noise level, gradient magnitude, and the resulting displacement budget); every parameter value above is then produced by combining a subset of those quantities with a simple analytic scale-matching relation. A few of them (the Farneb{\"a}ck polynomial smoothing, the Farneb{\"a}ck and Horn--Schunck iteration counts, and the FLCT window) involve no AIA quantity at all, and the FLCT threshold is driven by the noise floor of the noisier instrument in our wider pipeline rather than by AIA itself. The point of walking through both the analytic relations and the AIA-measured inputs together here is not to claim a uniquely optimal configuration, but to illustrate, with a concrete worked example, that this kind of reasoning can give each algorithm a sensible starting point instead of leaving it to be tuned blind.


\section{Velocity field estimation: 10 May 2022 X1.5 event}
\label{sec:demonstration}

We apply the FLCT, Horn--Schunck, Lucas--Kanade, and Farneb{\"a}ck algorithms to the 10 May 2022 dataset. Velocity fields are estimated from object masks generated by the Wavetrack framework on the AIA 193\,\AA{} running-difference images. For the purpose of pipeline integration, acceptable accuracy in correctly determining the shapes is one of the first priorities. The velocity magnitudes recovered by all four algorithms (Figs.~\ref{fig:fig_flct_panel}--\ref{fig:fig_fb}) are generally consistent with the wave-speed range reported by \cite{Rigney:2024}; the comparison below focuses on the spatial coherence of each velocity field and on the statistical agreement between the recovered speed distribution and the literature values.

\subsection*{FLCT velocity field on the X1.5 event}

\begin{figure}[!htb]
  \begin{center}
    \centering{\epsfig{file=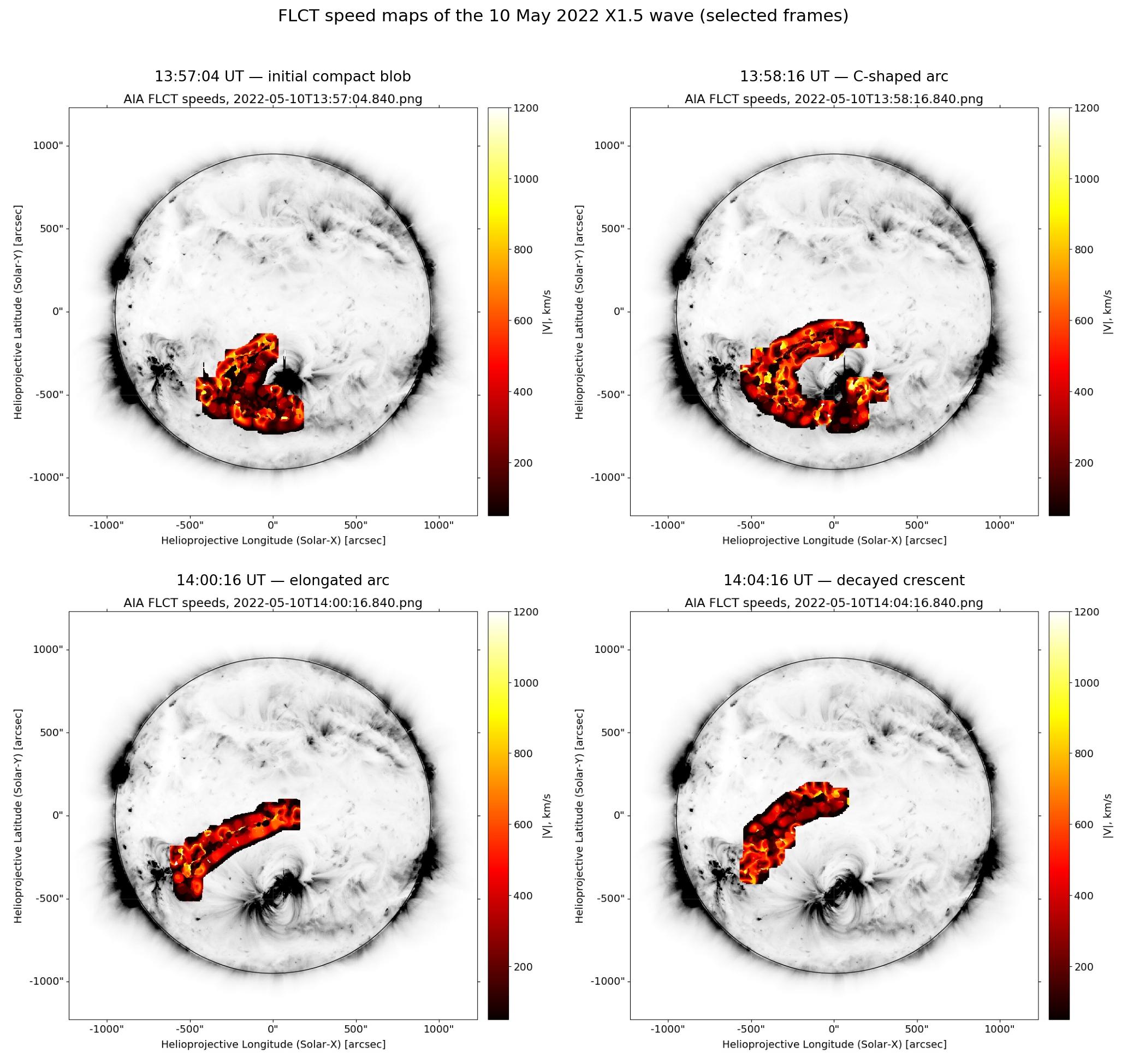, width=1.0\textwidth}}
    \caption{Selected FLCT speed maps of the 10 May 2022 X1.5 wave, computed on AIA 193\,\AA{} running-difference frames. The four panels capture the lifecycle of the coronal bright front: a compact blob over AR\,13006 at 13:57:04\,UT, a C-shaped horseshoe arc opening eastward at 13:58:16\,UT, an elongated arc rotated toward the disk centre at 14:00:16\,UT, and a decayed crescent at 14:04:16\,UT. In all four panels the rim and the southern/northern flanks (yellow, $\gtrsim$\,700\,km\,s$^{-1}$) move faster than the apex (red, $\sim$\,300--500\,km\,s$^{-1}$), consistent with an oblique-shock geometry.}
    \label{fig:fig_flct_panel}
  \end{center}
\end{figure}

\begin{figure}[!htb]
  \begin{center}
    \centering{\epsfig{file=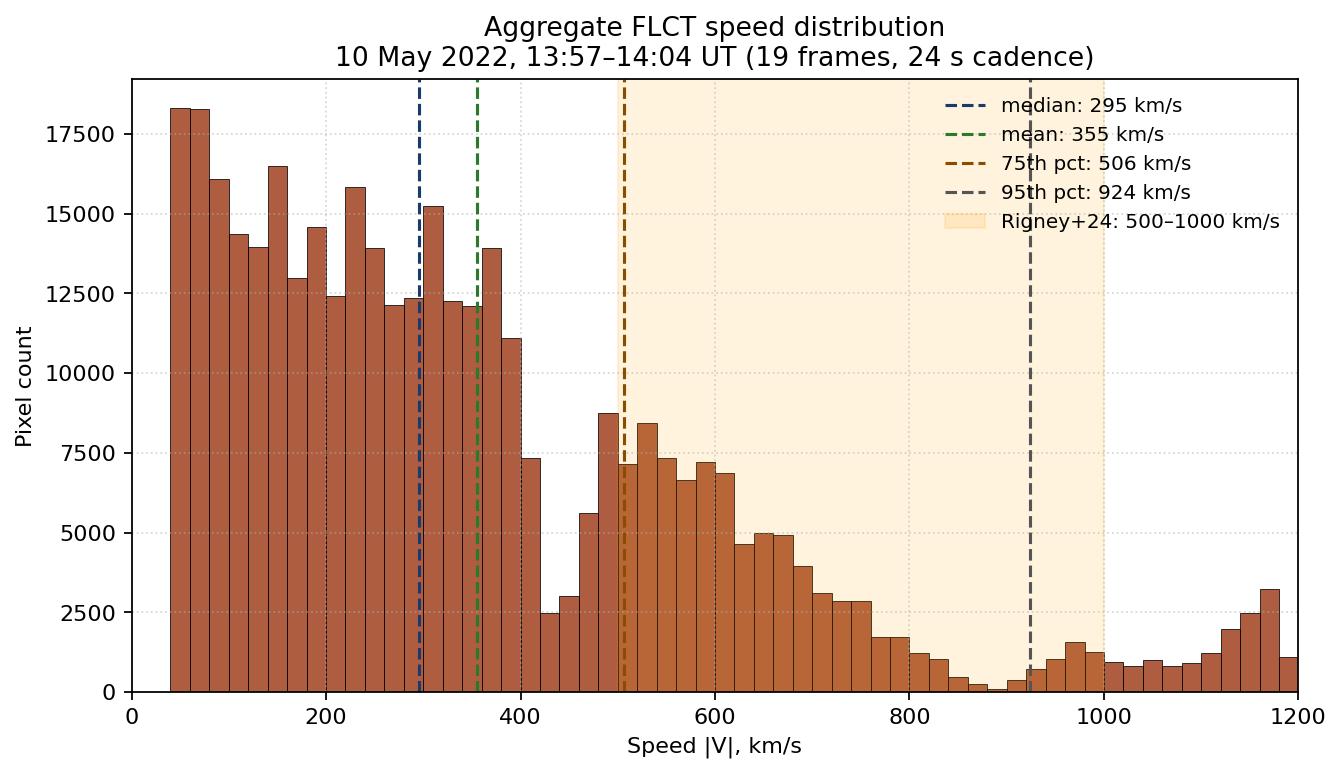, width=0.9\textwidth}}
    \caption{Aggregate distribution of FLCT speed magnitudes for the 10 May 2022 X1.5 event, pooled over all 19 frames between 13:57 and 14:04\,UT (380{,}595 valid pixels in total, 24\,s cadence). The distribution is bimodal: a dominant interior bulk between $\sim$50 and $\sim$400\,km\,s$^{-1}$ representing the bright-front body, and a secondary component between $\sim$500 and $\sim$700\,km\,s$^{-1}$ tracing the rim and the flanks. The high-speed tail extends to $\sim$1200\,km\,s$^{-1}$.
    The marked line indicates the $\sim$700\,km\,s$^{-1}$ FLCT-derived flank speed reported by \cite{Rigney:2024}, which falls within the secondary peak.
    }
    \label{fig:fig_flct_hist}
  \end{center}
\end{figure}

The FLCT velocity field (Fig.~\ref{fig:fig_flct_panel}) recovers the geometric evolution of the coronal bright front across the entire approx 7.5-minute sequence. At 13:57:04\,UT the wave appears as a compact, asymmetric blob anchored on AR\,13006, dominated by speeds in the 300--600\,km\,s$^{-1}$ range with isolated yellow pixels ($\gtrsim$\,800\,km\,s$^{-1}$) embedded in the core. Within $\sim$70\,s the front has expanded into a clear C-shaped arc opening eastward (13:58:16\,UT), with both flanks lit up in yellow while the apex remains in the red part of the colour scale. The arc continues to stretch and rotate, becoming elongated and tilted toward the disk centre by 14:00:16\,UT, and finally decays into a thinner crescent by 14:04:16\,UT in which only the leading southern flank still carries high-speed pixels. This visual narrative is qualitatively consistent with the ``flanks faster than apex'' behaviour expected for an obliquely propagating wave.

Quantitatively, the aggregate speed histogram (Fig.~\ref{fig:fig_flct_hist}) is markedly bimodal. The first lobe, between $\sim$50 and $\sim$400\,km\,s$^{-1}$, samples the interior pixels of the bright front and has a median speed of $\widetilde{|V|}=295$\,km\,s$^{-1}$ and a mean of $\langle|V|\rangle=355$\,km\,s$^{-1}$.
The second lobe, between $\sim$500 and $\sim$700\,km\,s$^{-1}$, samples the rim and the flanks; it extends up to the 75th-percentile speed of $506$\,km\,s$^{-1}$ and brackets the $\sim$700\,km\,s$^{-1}$ FLCT-derived flank speed reported by \cite{Rigney:2024} (marked in Fig.~\ref{fig:fig_flct_hist}). The 95th percentile reaches $924$\,km\,s$^{-1}$ and the tail extends to $\sim$1200\,km\,s$^{-1}$, exceeding the Rigney--24 FLCT-derived value; this tail could plausibly include a faster, shock-associated component, although we do not verify this independently here.

Examined frame by frame, the per-frame median is stable at $\sim$260--330\,km\,s$^{-1}$ while the 95th percentile decreases monotonically from $\sim$980\,km\,s$^{-1}$ early in the sequence to $\sim$700--800\,km\,s$^{-1}$ towards the end, exposing the deceleration of the front already visible as the fading high-speed tail in the speed-vs-time spectrogram of the same event.

The statistical agreement with \cite{Rigney:2024} is broadly consistent with both the FLCT outputs and the calibration approach of Sec.~\ref{sec:fitting}: the noise-to-contrast ratio measured on the AIA frames, $\eta=\sigma_n/\Delta I\approx 0.09$, is about a factor of three below the LASCO-derived correlation floor $\tau=0.3$ used here, so the threshold is conservative for AIA in absolute terms, and we did not visually identify spurious high-speed pixels in the quiet-corona regions of Fig.~\ref{fig:fig_flct_panel} (though this was not quantified). Conversely, the choice of the Gaussian weighting kernel (mandatory for FLCT's FFT-based correlation, to suppress wrap-around artefacts at the patch edges) is what allows the rim pixels to be tracked despite their proximity to the bright-front boundary; without the Gaussian taper, the correlation peaks at the rim would be biased toward the static background. The FLCT result of Fig.~\ref{fig:fig_flct_panel} therefore illustrates the same parameter-scaling logic discussed for the HS/LK/FB outputs below, applied here to a Fourier-correlation framework.

\subsection*{Optical-flow velocity fields on the X1.5 event}

\begin{figure}[!htb]
  \begin{center}
    \centering{\epsfig{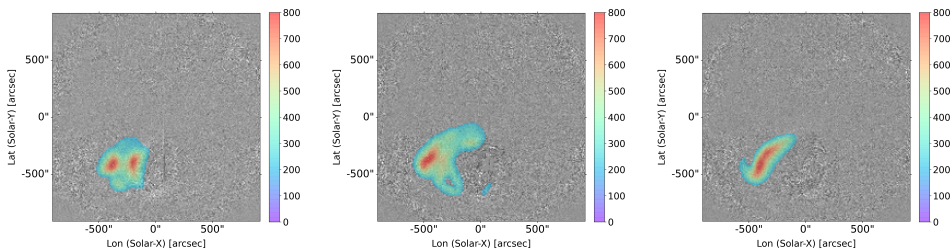}}
    \caption{Shock-wave propagation calculated using the Horn--Schunck optical-flow algorithm with the parameters derived in Sec.~\ref{sec:fitting}. The sequence shows coherent propagation, with a smooth and continuous shock-wave advancement over time in this example.}
    \label{fig:fig_hs}
  \end{center}
\end{figure}

\begin{figure}[!htb]
  \begin{center}
    \centering{\epsfig{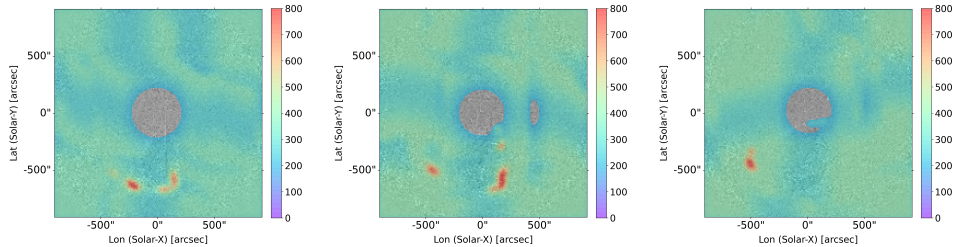}}
    \caption{Shock-wave propagation calculated using the Lucas--Kanade optical-flow algorithm with the parameters derived in Sec.~\ref{sec:fitting}. The sequence depicts a more fragmented and less consistent wave propagation.}
    \label{fig:fig_lk}
  \end{center}
\end{figure}

\begin{figure}[!htb]
  \begin{center}
    \centering{\epsfig{file=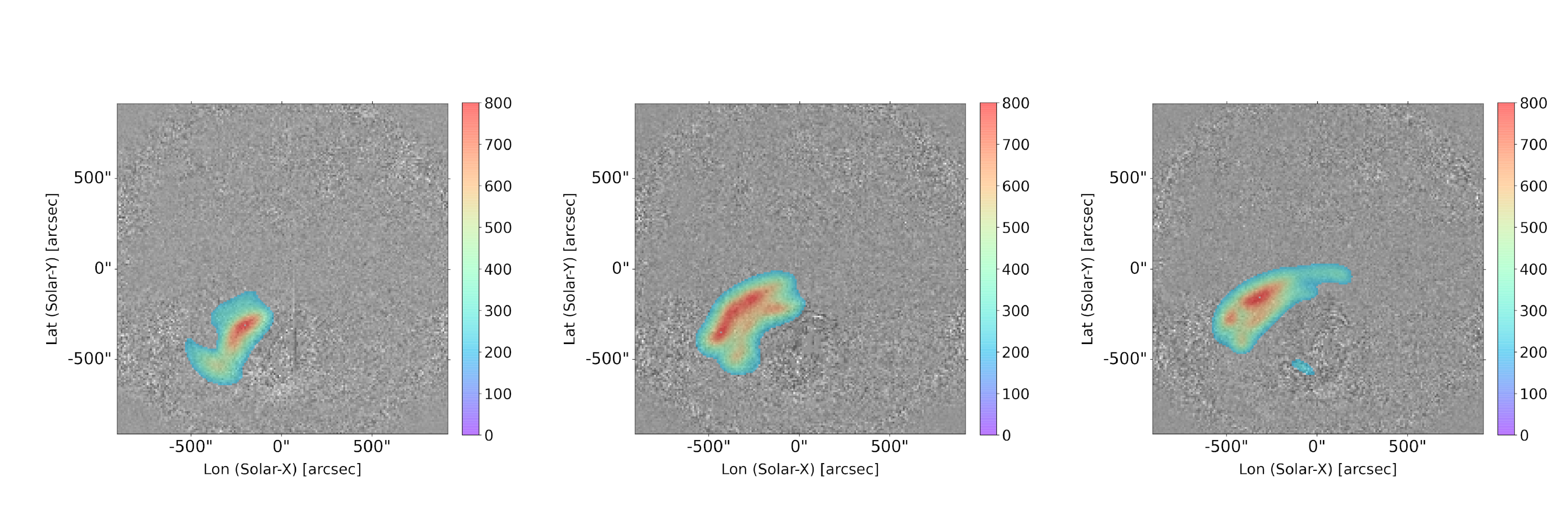, width=1.0\textwidth}}
    \caption{Shock-wave propagation calculated using the Farneb{\"a}ck optical-flow algorithm with the parameters derived in Sec.~\ref{sec:fitting}. The images show a well-defined shock-wave progression, comparable in coherence to the Horn--Schunck algorithm in this example, with a continuous wavefront.}
    \label{fig:fig_fb}
  \end{center}
\end{figure}

The Horn--Schunck algorithm (Fig.~\ref{fig:fig_hs}) appears to resolve the shock wave's propagation reasonably well in this example, providing a smooth, continuous wavefront. This is consistent with the parameter choice discussed in Sec.~\ref{sec:fitting}: the regularization $\alpha = 0.3$--$1.0$ was set so that the diffusion length $\ell_\mathrm{HS} \approx \alpha/(\|\nabla I\|/S)$, with $S \approx 1\times10^{4}\,\mathrm{DN\,s^{-1}}$ as in Sec.~\ref{sec:fitting}, matches the feature scale, and the 500 Jacobi iterations are intended to be enough to propagate information across the feature radius. The result is a globally coherent field, consistent with what Horn--Schunck is designed to enforce; we did not notice obvious artefacts or discontinuities in this example.

Previously we had mixed results on the precision of the Lucas--Kanade algorithm \cite{Stepanyuk:2024} applied to ground-based and remote-instrument data. For this event, the algorithm (Fig.~\ref{fig:fig_lk}) exhibits notably poorer performance in this example: the shock-wave propagation appears fragmented, with notable discontinuities and inconsistencies in wavefront shapes across the sequence. This outcome is also consistent with the parameter scaling derived above. The Lucas--Kanade window $W = 7$, set by $W \approx w/2$ with the AIA-measured rim width $w = 11$\,px, resolves the rim edge but is much smaller than the feature radius $r \approx 15$--22\,px. LK therefore enforces \emph{only} local coherence and cannot, by construction, propagate flow information across regions of low gradient between rim segments. The local window-based approximation also struggles with the regions of large displacement implied by $\Delta \approx 14$--28\,px per 12-second cadence -- larger than the window itself.

The Farneb{\"a}ck algorithm (Fig.~\ref{fig:fig_fb}) performs comparably well to Horn--Schunck in this example. The shock waves appear well defined and maintain continuity and coherence across frames. The pyramid scale $s = 0.3$ at $L=1$ gives an effective window of $\sim 50$\,px, which is intended to be large enough to absorb the AIA-measured displacement budget $\Delta_\mathrm{max} \approx 27.6$\,px and allow the polynomial expansion of Eq.~\eqref{eq:farneback} to capture the nonlinear intensity variations across the front. Farneb{\"a}ck therefore combines the local accuracy of a polynomial fit with the longer-range tracking ability of a multi-level pyramid -- representing the shock-wave propagation in this example without the global oversmoothing risk that can affect Horn--Schunck-type regularization, or the local-window limitation seen in Lucas--Kanade here.

In summary, the scale-matching arguments of Sec.~\ref{sec:fitting} are consistent with the relative performance of the four algorithms observed on the X1.5 event: FLCT, Horn--Schunck and Farneb{\"a}ck all possess a mechanism (Fourier patch correlation with a Gaussian taper, global regularization, pyramid coverage) to bridge the displacement budget $\Delta_\mathrm{max} \gg W$ in the AIA data, whereas Lucas--Kanade does not.
FLCT additionally returns a speed distribution whose secondary peak (500--700\,km\,s$^{-1}$) brackets the $\sim$700\,km\,s$^{-1}$ FLCT-derived flank speed of \cite{Rigney:2024}, while its 95th percentile ($924$\,km\,s$^{-1}$) exceeds it, plausibly tracing a faster shock-associated component; this is at least qualitatively consistent with the calibration used here, though it remains a single-event comparison rather than an independent validation.
The Lucas--Kanade algorithm's local method, by contrast, appears less practical in this configuration for on-disk shock-wave propagation than Horn--Schunck's global regularization, Farneb{\"a}ck's polynomial-pyramid approach, or FLCT's Fourier correlation, which in this example produced visually clearer and more spatially coherent results.


\section {Limitations and Outlook for Future Work}

While the present study provides a conceptual framework and a transparent, physics-informed methodology for tuning optical-flow and local-correlation algorithms to moderately sized coronal bright fronts, two important limitations should be noted. First, both the parameter derivation above and the performance demonstration rely on a single, well-observed X1.5-class flare event (10 May 2022). Although this event is scientifically rich and representative of complex on-disk shock-wave propagation, the recommended parameter set has not yet been tested across a statistically meaningful sample of events. Its generalizability to quieter EUV waves, limb or off-disk geometries, lower signal-to-noise regimes, or data from other instruments (e.g., LASCO C2/C3 where the conservative correlation threshold $\tau$ = 0.3 was chosen) therefore remains an open question. The provided estimates are order-of-magnitude constraints rather than exact derivations.

These limitations point directly to several follow-up studies that would strengthen and extend the current concept. A natural next step is to perform a multi-event validation campaign using a carefully curated catalog of EUV waves (e.g., 20–50 events spanning quiet-Sun, active-region, and limb cases across multiple solar cycles and instruments). This would allow statistical assessment of parameter robustness and, if needed, the development of adaptive or event-specific tuning rules.

Quantitative metrics—such as velocity error relative to manual tracking, wavefront continuity scores, divergence/curl statistics, synthetic-data benchmarks, or cross-comparison with independent kinematic methods—on a larger datasets would be required for applications in pipelines.

\section*{Conclusion}
Tuning of velocity estimation algorithms depends strongly on the spatial scale and motion intensity of the structures of interest. FLCT, and to some extent Lucas-Kanade when properly windowed and thresholded, appear effective in this example for moderate-scale coherent motions. Farneback provides dense and efficient estimations but requires careful adjustment of smoothing parameters to avoid over-generalization. Horn-Schunck appears reasonably robust in this example but requires balancing between structure-preserving and magnitude-sensitive regimes.

We discussed a practical set of algorithm parameters—arrived at by combining analytic reasoning with empirical measurements—and illustrated the performance of velocity-field estimation techniques for on-disk shock-wave propagation during the X1.5-class event of 10 May 2022. In practical applications, adaptive parameter-selection strategies are likely to deliver the most consistent results across varying motion fields.

\section*{Acknowledgments}
This work is supported by the Bulgarian National Science Fund, VIHREN program, under contract KP-06-DV-8/18.12.2019 (MOSAIICS project)

\bibliographystyle{apalike}
\bibliography{unet_aia_wavetrack}

\end{document}